# Self-assembled nanostructured metamaterials

Virginie Ponsinet[1], Alexandre Baron[1], Emilie Pouget[2], Yutaka Okazaki[2], Reiko Oda[2] and Philippe Barois[1(a)]

[1] *Centre de Recherche Paul Pascal, Bordeaux University – CNRS, 115 Avenue Albert Schweitzer, 33600, Pessac, France*
[2] *Institut de Chimie et Biologie des Membranes et des Nano-objets, Bordeaux University – CNRS, 2 rue Robert Escarpit, 33607 Pessac, France*

**Abstract** – The concept of metamaterials emerged in years 2000 with the achievement of artificial structures enabling non conventional propagation of electromagnetic waves, such as negative phase velocity of negative refraction. The electromagnetic response of metamaterials is generally based on the presence of optically-resonant elements – or meta-atoms – of sub-wavelength size and well designed morphology so as to provide the desired electric and magnetic optical properties. Top-down technologies based on lithography techniques have been intensively used to fabricate a variety of efficient electric and magnetic resonators operating from microwave to visible light frequencies. However, the technological limits of the top-down approach are reached in visible light where a huge number of nanometre sized elements is required. We show here that the bottom-up fabrication route based on the combination of nanochemistry and of the self-assembly methods of colloidal physics provide an excellent alternative for the large scale synthesis of complex meta-atoms, as well as for the fabrication of 2D and 3D samples exhibiting meta-properties in visible light.

**Introduction.** – Metamaterials are artificial composite materials that exhibit extreme physical properties that do not exist in natural materials. The origin of the scientific field is often associated with the theoretical work of Veselago in 1968 who discussed the propagation of light in a hypothetical negative index material (NIM) resulting from simultaneous negative values of the electric permittivity $\varepsilon$ and of the magnetic permeability $\mu$ [1], but the interest for metamaterials really started and grew tremendously at the start of the years 2000 when the first realization of a NIM occurred, based on the novel design proposed by Sir John Pendry of a sub-wavelength inductive-capacitive resonant circuit (split ring resonator or SRR) enabling the control of the magnetic response to light [2,3]. Although limited to microwave frequencies in the GHz range, the demonstration of the non-natural NIM property suggested that unprecedented extraordinary optical applications such as perfect lensing [4] or cloaking [5] were achievable, especially since the recipe for the design of metamaterials was given: the functionality is based on resonant inclusions of sub-wavelength size (meta-atoms) distributed in a transparent matrix. The resonant behaviour of the inclusions warrants an intense optical response upon approaching the resonance frequency $\omega_0$ (high $\varepsilon$, high $\mu$) as well as a phase shift leading to the desired sign inversion upon crossing $\omega_0$. The sub-wavelength condition is required to enable homogenization, i.e. a valid description of the optical properties by the usual effective parameters $\varepsilon$ and $\mu$ or equivalently the refractive index $N = (\varepsilon\mu)^{1/2}$ and the optical impedance $Z = (\mu/\varepsilon)^{1/2}$. The concept of metamaterials, which can be summarized as the full control of the reflection, refraction, phase and polarization of light waves at the sub-wavelength scale, hence led to the blossoming of a tremendous set of potential applications that were abundantly discussed in the literature like filters, light couplers, light concentrators, field enhancers for sensing or non-linear optics, miniaturized antennas to name a few [6]. The concept was actually extended to all types of waves with spectacular applications in acoustics [7], mechanics [8,9], ocean [10,11] and seismic waves [12].

Coming back to electromagnetic waves, a vast majority of the experimental realizations are devices fabricated by lithography techniques. This top-down approach offers a quasi-perfect control of two dimensional composite structures, sometimes referred to as metasurfaces. It is however hard to extend to 3D materials, because the number of inscribed resonators is generally limited and the lithography techniques reach their limits when nanometer resolution is needed for visible light applications. In this case, the size of the resonators and the distance between them should not exceed ~100 nm to satisfy the homogeneity condition, which requires ~$10^{12}$ resonators per mm$^3$ to make a bulk meta-material or $10^8$/mm² for a metasurface. The bottom-up approach that combines nano-chemistry for the synthesis of optical resonators and colloidal physics for their self-assembly in 2D or 3D structures appears as an appealing alternative for the fabrication of metamaterials operating in visible light or near IR [13,14]. It can easily produce and handle large volumes of materials at low cost, though at the expense of a lower degree of structural control of the nanostructure. The aim of this paper is to illustrate the bottom-up approach with a selection of recent examples of self-assembled metamaterials.

**Self-assembly of resonators** – Advanced optical functionalities require resonators of complex morphologies that can be achieved by self-assembly of simpler units. The power of the bottom-up approach lies in its ability to synthesize a huge number of resonators ($\sim 10^{13}$ per batch in the laboratory) which can be subsequently assembled into macroscopic materials. Two examples are given below.

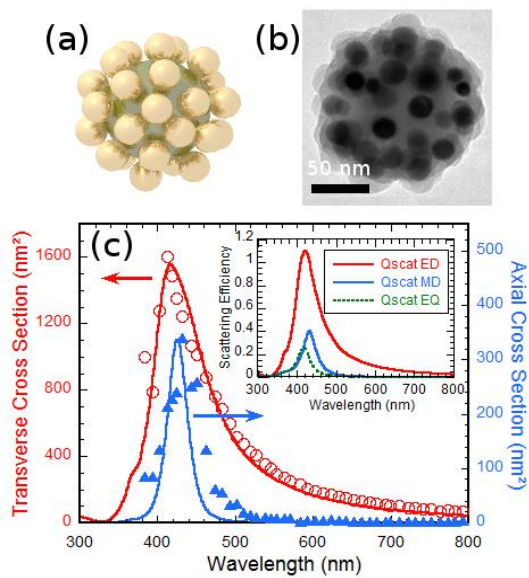

been achieved in devices based on pairs of plasmonic elements organized on a surface by the top-down approach [16-19]. However, the optical magnetic response of such metasurfaces is anisotropic and highly non-local. It is depicted by spatial dispersion of the dielectric permittivity rather than by a homogeneous permeability parameter $\mu$ [20,21]. An alternative model of raspberry-like isotropic magnetic nanoclusters (MNC) shown in Fig. 1a was proposed by Simovski and Tretyakov [22] in which the light wave induces loops of resonant plasmonic currents [23]. Fig. 1b shows a realization of the MNC by Gomez-Graña *et al.* [24] formed by silver nanoparticles surrounding a spherical silica core. The spontaneous self-assembly of the raspberry structure results from electrostatic interaction between the positively charged core and the negatively charged silver satellites, the final MNC being encapsulated in a thin silica layer to warrant its structural stability. The magnetic response of the MNCs in visible light was demonstrated by analyzing the polarization of the light scattered by a dispersion of MNCs in water. Fig. 1c shows plots of the electric and magnetic scattering cross sections, which are perfectly consistent with numerical simulations [24].

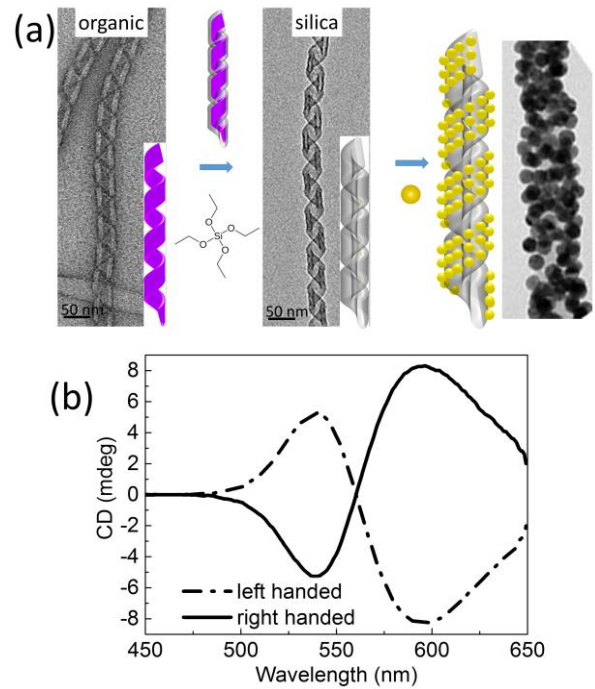

Fig. 1: (a) Simovski-Tretyakov model of raspberry-like magnetic nanoclusters. Metallic nanoparticles are distributed around a dielectric core. (b) Electron micrograph of a plasmonic raspberry synthesized by self-assembling silver satellites (25 nm) around a silica core (100nm). (c) Electric (red) and magnetic (blue) scattering cross-sections of the magnetic nanoclusters. Solid lines are numerical simulations. Inset shows the magnetic dipole contribution (MD) with the electric dipole (ED) and quadrupole (EQ) [24].

*Magnetic nanoclusters.* The generation of optical magnetism in visible light is a tough challenge: the response time of para- and ferromagnetism is far too slow and natural diamagnetism is notoriously negligible [15]. Nonetheless, the generation of intense magnetic dipoles at the frequency of visible light has

Fig. 2: (a) Synthesis of the chiral meta-atoms: from organic self-assembly forming nanohelices used as template for the silica transcription and grafting of 9nm GNPs on their surfaces. (b) Circular dichroism spectra showing the opposite signals for left- and right-handed GNPs/silica helices [26].

*Plasmonic nanohelices*. Enhancing the chirality of nanoobjects or substrates is another important challenge of metamaterials with applications in chiral recognition, sub-wavelength circular polarisers or low index materials. A natural strategy for the synthesis of super-chiral meta-atoms consists of organizing nano-resonators in a helical fashion [25]. Cheng *et al.* used chiral colloids as templates for grafting gold nanoparticles (GNPs) [26]. These hybrid nano-helices are synthesized from organic chiral self-assemblies forming very well defined helices or ribbon structures (Fig. 2a) [27]. The mineralization of these organic self-assemblies enables the creation of silica nano-helices with very well controlled morphologies in terms of diameter and pitch (typical values are, for the helices diameters of 35nm and pitches of 65nm and for twisted ribbons widths of 20nm and pitches of 95nm). The handedness of the structures is governed by the constituent molecular chirality [28-30]. These nanometric silica helices are then used as chiral templates in order to organize GNPs to prepare a collection of helical GNP superstructures. Varying the GNP size between 4 nm and 10 nm, as well as the surface chemistry of both the GNPs and the silica nano-helices, it is possible to span a large range of grafting densities. When the GNPs are densely adsorbed on the helices, they follow the helicity of the silica template and strong chiroptical activity is evidenced by electronic circular dichroism spectroscopy at the wavelength of the surface plasmon resonance (SPR) of the GNPs, showing opposite signal for left handed and right handed helices (Fig. 2b). The measured anisotropy factor (g-factor) is of the order of $10^{-4}$, *i.e.* ten-fold larger than typical values reported in the literature. The optical activities depend on the size and the organization of the GNPs on the surface of the helical silica template which are closely correlated with the surface chemistry of both the GNPs and the silica helices [26,31].

**Self-assembly of a macroscopic material.** – Once the meta-atoms have been synthesized, the next challenge is to assemble them into a bulk material or into a metasurface while preserving their optical functionality. In contrast with photonic crystals, which operate in the diffractive regime, where the structural ordering is critical, metamaterials are not much affected by structural disorder, as long as the response of the meta-atoms is not degraded by their mutual coupling, because they function in the regime of homogeneous optical response. The popular self-assembly routes of colloidal physics hence appear as excellent tools for the macroscopic organisation of large numbers of meta-atoms. Thin films of large area can for example be produced by Langmuir-Blodgett assembly, dip- and spin-coating or meniscus evaporation. 3D materials may result from sedimentation of a suspension of meta-atoms or controlled evaporation of their solvent. Two examples are illustrated below.

*Bulk magnetic metamaterial*. The magnetic nanoclusters shown in Fig. 1b have been assembled in a bulk 3D material in a micro-evaporator made in a microfluidic chip designed so as to enable a slow evaporation of water across a thin polymer (PDMS) membrane [32]. After complete evaporation, a dense material made of a random close-packing of MNCs is formed which replicates the shape of the microfluidic canal (Fig. 3a). The optically flat surface and the large thickness of the metamaterial enables a full determination of the optical parameters $\varepsilon$ and $\mu$ by variable angle spectroscopic ellipsometry. The magnetic permeability $\mu$ is shown (Fig. 3b). Two results of this study are remarkable. First, the permeability deviates significantly from 1. Variations from 0.8 to 1.45 correspond to a magnetic susceptibility $\chi=\mu-1$ ranging from -0.2 to 0.45, three to five orders of magnitude higher than natural diamagnetism. Second, the angular study shows that the values of the optical parameters $\varepsilon$ and $\mu$ do not depend on the angle of incidence.

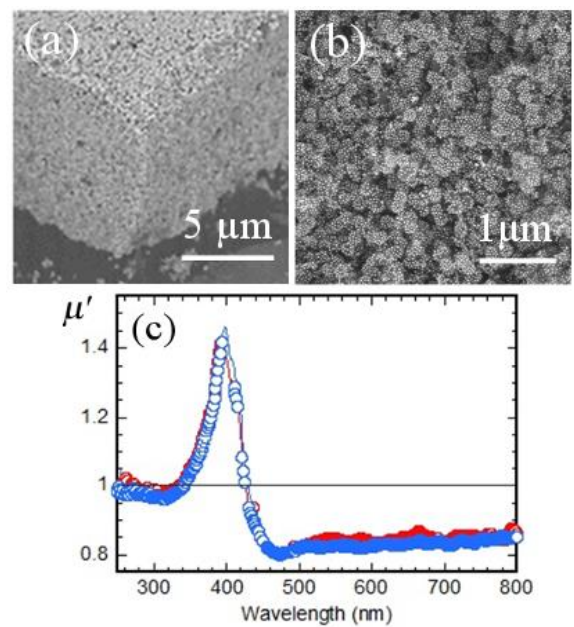

Fig. 3: (a,b) SEM views of the assembled magnetic metamaterial showing flat surfaces in (a) and random packing in (b). (c) Real part of the magnetic permeability extracted from spectroscopic ellipsometry. Red and blue colours correspond to two different samples. Natural materials lie on the black line at $\mu'=1$

This example illustrates two positive points of the bottom-up approach. The sample thickness larger than the absorption length equivalent to a semi-infinite geometry enables an analytical extraction of the optical parameter without any *ad-hoc* model. Moreover, the random packing of the MNCs implies a continuous rotational symmetry of the nanostructure which minimizes the effects of spatial dispersion as explained in reference 24.

*Resonant metasurfaces*. The optical properties of 2-D arrays of resonators have been widely investigated by theoretical, numerical and experimental studies. Malassis *et al.* prepared dense arrays of silver nanospheres of large area by transferring Langmuir monolayers on a flat surface [33, 34].

Interestingly, the distance (and hence the electromagnetic coupling) between the silver resonators can be controlled by encapsulating them in a silica layer of controlled thickness. Fig. 4b-c shows a film made of 6 layers of core-shell silver@silica on a silicon substrate [33] by 6 successive Langmuir-Schaefer transfers, a variant of the Langmuir-Blodgett technology in which the Langmuir monolayer is transferred to an immersed substrate by lowering the water level of a Langmuir trough. The presence of sharp, blue-shifted optical resonances (with respect to single resonators) was demonstrated in these films which exhibit the interesting property of topological darkness, or plasmonic Brewster extinction whereby the reflection from an absorbing substrate is totally extinguished by the optical interference in the film [34].

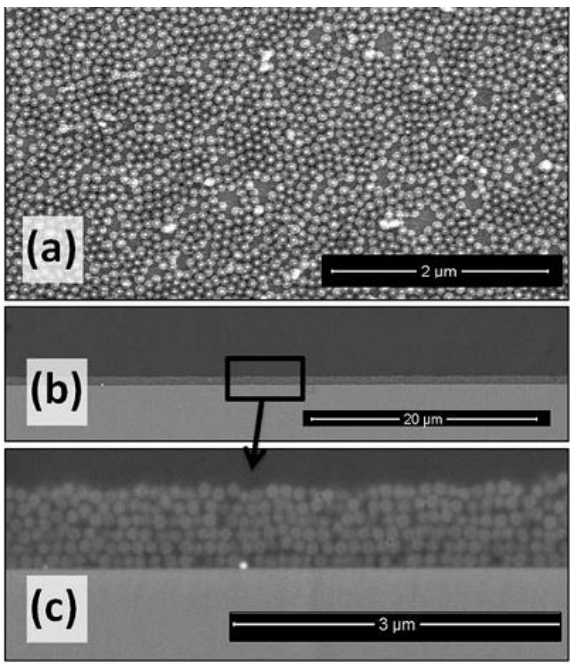

Fig. 4: (a) SEM top view of a monolayer of core-shell Ag@SiO$_2$ nanoparticles transferred on a silicon substrate by the Langmuir-Schaefer technique. (b, c) Side views of a film made by 6 successive transfers (from [32]).

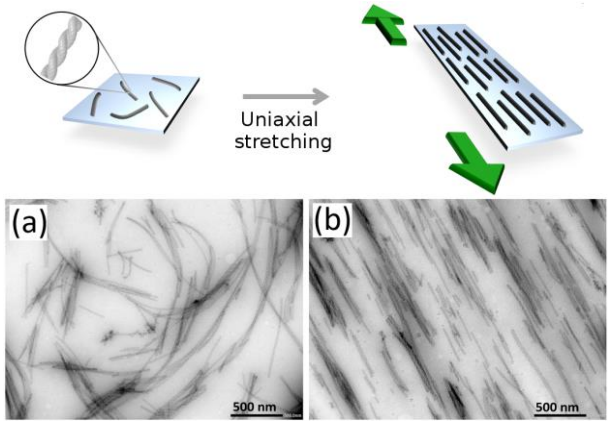

Fig. 5: Silica nano-ribbons dispersed in a poly(ethylene-co-vinyl acetate) polymer matrix (a) randomly oriented and (b) well aligned after unidirectional stretching

*Chiral films*. Organizing elongated objects like nanorods or nanohelices within a thin film usually requires the control of the orientation as an additional degree of freedom. Several fabrication techniques promote the orientation of anisotropic objects on a surface like dip-coating [35], grazing incidence spraying [36] or surface patterning [37] to name a few recent examples. Another approach is illustrated in Fig. 5 in which the silica nanohelices mentioned above are incorporated in a polymer film which is subsequently stretched along one direction. Fig. 5a shows bare silica ribbons with random orientation after dispersion in a polymer matrix of poly(ethylene-co-vinyl acetate). By stretching 10 times the matrix, the transparent film becomes slightly hazy as the chiral nanoribbons get perfectly oriented along the strain direction Fig 5b).

**In-situ synthesis within a self-assembled template.** – Another self-assembly route for the fabrication of metamaterials consists in introducing resonators in selected regions of a pre-existing self-assembled nanostructure, which acts as a template. An example is given below.

*Self assembled hyperbolic metamaterials*. Some meta-properties are related to the anisotropy of the materials rather than to the presence of complex resonant objects. This is the case for hyperbolic metamaterials, in which a combination of nanostructuration and anisotropy gives rise to non-natural dispersion relations, and which are now regarded as highly promising metamaterials, because of their ability to provide a multi-functional platform to reach different meta-properties [38].

A hyperbolic metamaterial presents two components of the dielectric permittivity tensor $\underline{\varepsilon}$ with opposite signs, as if it behaves like a metal ($\varepsilon_i<0$) along at least one direction and like a dielectric ($\varepsilon_j>0$) along at least another. Consequently, the isofrequency dispersion relation exhibits a hyperboloid branch, which notably allows for the propagation of large magnitude wavevectors. These large wavevectors carry the information of fine details and are usually associated to evanescent waves in natural materials, which make them inaccessible to regular microscopes. Soft matter self-assembly can provide powerful fabrication routes for such extreme anisotropy. Liquid crystals and block copolymers, in particular exhibit self-assembly properties described theoretically and experimentally to a high degree of detail for the last fifty years [39, 40]. They present spontaneous molecular organization, possibly strongly anisotropic, with long-range order and characteristic sizes in the range 2-500 nm. These are, however, organic materials, with moderate susceptibilities and low optical constant contrasts. Therefore, in their native state, they should be considered essentially as 'optically neutral' templates whose role is to spatially organize optically resonating nanoparticles. Block copolymer self-assembled nanostructures can be used to template and order solid nanoparticles [41, 42]. Such nanoparticle ordering into complex hybrid nanostructures have proven useful to monitor optically probed nanoscale sensing [43] or produce photonic bandgap materials [44].

Lamellar nanocomposite thin films have been produced on a silicon substrate via the *in situ* synthesis of gold nanoparticles, selectively grown in the poly(2-vinylpyrridine) domains of a lamellar phase of poly(styrene)-b-poly(2-vinylpyrridine) block copolymer. The SEM images in Fig. 6 show that the films are structurally uniaxial and homogeneous, enabling the definition of the dielectric permittivity tensor with ordinary (parallel to the substrate, $\varepsilon_{//}$) and extraordinary (normal to the substrate, $\varepsilon_\perp$) components. Using variable angle spectroscopic ellipsometry, a very large permittivity anisotropy has been measured, thus reaching appropriate conditions for hyperbolic properties, with $\varepsilon_{//}<0<\varepsilon_\perp$ in a wavelength domain close to the plasmon resonance of the gold nanoparticles loaded in the nanocomposite (Fig. 6) [45].

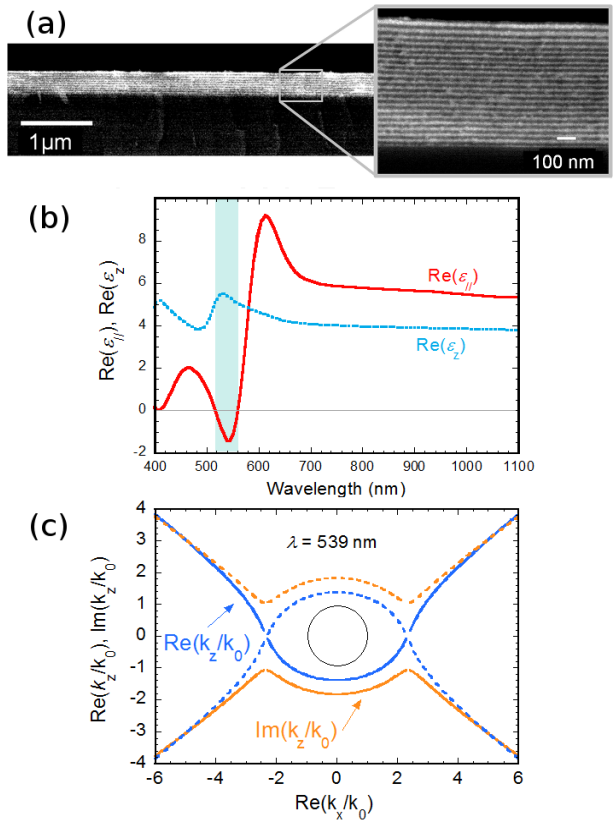

Fig. 6: (a) Backscattering SEM side view of a 680nm-thick film of alternating layers of pure polymer (appearing black) and of gold:polymer nanocomposite (NC, appearing white). Inset: Zoom of the same film (bar=200nm). (b) Parallel (red) and perpendicular (blue) components of the uniaxial dielectric function, extracted from spectroscopic ellipsometry, of a 265nm-thick film for a gold volume fraction $f_{Au} \sim 30\%$ in the NC layers. The shaded area highlights the hyperbolic regime. (c) Hyperbolic dispersion relation calculated in at $\lambda$=539 nm for $k_y = 0$, showing the real (blue) and imaginary (orange) parts of the $k_z$ component of the wave-vectors in the material as a function of the $k_x$ component. Continuous and dashed lines correspond to opposite solutions. The thin black circle shown for reference corresponds to wavevectors propagating in air $k_x^2 + k_z^2 = k_o^2$ ,.

In the hyperbolic region, the dispersion relation allows propagative modes for large values of $|k_x|$, potentially providing super-resolution. These modes are however significantly attenuated due to the resonant – and therefore lossy – nature of the permittivity. These detrimental losses need to be minimized, which should be done by optimizing the quality factor of the resonances, through tuning the nature, size and organization of the nanoparticles.

**Perspectives.** – Just like their top-down counterparts, bottom-up metamaterials are severely affected by optical losses. Losses are indeed fundamentally associated with highly dispersive resonances by the principle of causality. The compensation of plasmonic losses by an active gain medium has been proposed as a possible solution in which energy is injected at the plasmon frequency by external sources via optically pumped fluorophores or quantum dots [46]. This active technology, still to be developed, is likely to be expensive and limited to high added-value applications.

In passive systems, two options can be explored to reduce the effect of losses in future bottom-up metamaterials. The first one consists in exploring all-dielectric metamaterials which exhibit much lower losses than plasmonic materials [47], the other option consists in producing metasurfaces rather than metamaterials [6].

*All-dielectric metamaterials*. A dielectric nanoparticle with an index of refraction *n* exhibits Mie resonances when the typical size of the nanoparticle *a* is on the order of the wavelength *λ* of the impinging radiation. The scattering properties of these nanoparticles are perfectly described by Mie scattering theory [48] and two strong resonances of interest are the so-called magnetic-dipole resonance, which occurs when $a \sim \lambda/(2n)$ and the electric-dipole resonance, which occurs when $a \sim \lambda/n$. The larger the index of refraction of the material, the smaller the nanoparticle can be for a given wavelength of operation, therefore allowing for the homogeneization condition $a \ll \lambda$. Semiconductors such as silicon are good candidates. Indeed, the index of refraction of crystalline silicon (Si-c) is larger than 3 across most of the near infrared spectrum and even rises above 4 in the visible range. Several commercially available particles of spherical shape that exhibit large values of *n*, such as amorphous silicon (Si-a) and crystalline $TiO_2$, can be purchased at low cost and in large quantities in the form of powders, which can subsequently be suspended in a solvent so as to obtain colloidal dispersions of all-dielectric optical resonators.

Fig. 7a provides the typical spectral variations of the scattering cross section efficiency of a silicon nanoparticle of 150 nm in diameter. We see that the total scattering ($\sigma$) is the sum of the scattering from the magnetic and electric dipoles. The magnetic dipole moment is almost one order of magnitude larger than that of raspberry-shaped plasmonic magnetic nanoclusters [24,49], and it is moreover comparable in amplitude to the electric dipole. Following extended Maxwell-Garnett theory [50] we can infer the effective medium properties of a host medium of dielectric constant $\varepsilon_h$ containing such spherical inclusions.

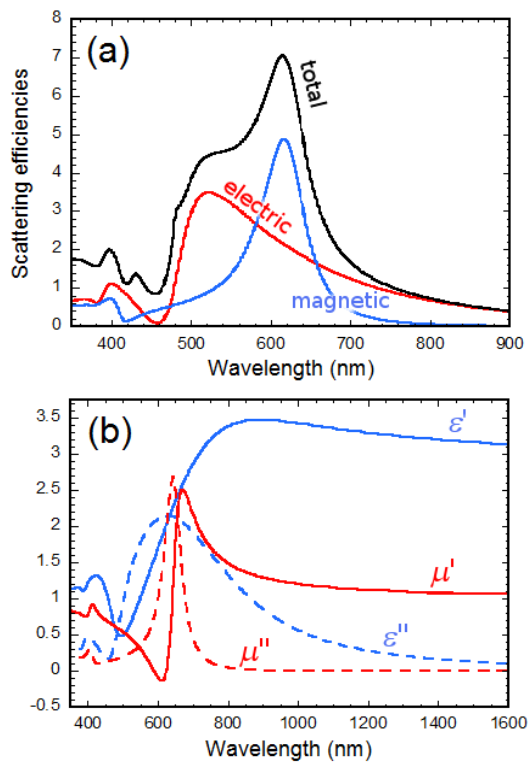

**Fig. 7:** Optical properties of a self-assembled all-dielectric metamaterial. (a) Spectral variations of the total scattering efficiency (black curve) computed for a 150 nm silicon nanoparticle dominated by the scattering of a magnetic dipole (blue curve) and an electric dipole (red curve). (b) Spectral variations, based on an extended Maxwell-Garnett mixing law, of the complex electric permittivity $\varepsilon = \varepsilon' + i\varepsilon''$ and magnetic permeability $\mu = \mu' + i\mu''$ of a three-dimensional metamaterial composed of the same silicon nanoparticles (60% volume fraction).

Fig. 7b shows the electric permittivity $\varepsilon$ and magnetic permeability $\mu$ computed for a volume fraction of 60% close to that of typical dense metamaterials self-assembled by microfluidic evaporation [24,51]. Both $\varepsilon$ and $\mu$ show large spectral variations in the visible and large and comparable magnitudes. Notably for $\lambda \sim 658$ nm, the electric and magnetic dipoles are in phase and have equal amplitudes (see Fig. 7a), such that light is scattered in the forward direction following the well-known Kerker condition [52]. Consequently, $\varepsilon \sim \mu$ for the metamaterial (Fig. 1b) and the impedance of the air/metamaterial interface is $z = (\mu/\varepsilon)^{1/2} \sim 1$ which corresponds to a cancellation of the reflection. This means that high-index all-dielectric metamaterial/air interface acts as a perfect coupler of light to the metamaterial, which is of particular interest for the enhancement of light-matter interaction as well as perfect absorption.

*Self-assembled metasurfaces*. Another aim in the self-assembly of optical resonators is the production of two-dimensional metamaterials. Since the skin depth rarely exceeds a single-layer of resonators in resonant metamaterials, the functionality can be preserved with a so-called metasurface, defined as a two-dimensional artificial material that provides some optical functionality. Notably, most metasurfaces up to date have been designed with the aim of achieving spatial control of the wavefront. By carefully arranging resonators, each of which will scatter light with a tailored phase-shift with respect to the impinging wave, one can design a surface that effectively scatters as a bulk medium shaped so as to imprint a given phase. Using this principle, many flat-optical components have been imagined, such as flat lenses, axicons, beam steerers or diffractive elements [6].

**Conclusion.** Bottom-up fabrication can certainly play a significant role in the area of metamaterials and metasurfaces for applications where a perfect ordering of the meta-atoms is not necessary. Of course, producing amorphous materials is what comes to mind first, which would enable a cost-effective production of large-area devices. The bottom-up approach is particularly suited to hierarchical self-assembly, which makes it possible to design the nanoresonator independently from the assembled material. Moreover, chemical synthesis is able to produce huge amounts of meta-atoms in a single batch (e.g. > $10^{13}$ in references [24, 33, 49]), far beyond the capabilities of the lithography methods. As a result, it should be possible in principle to produce inks containing meta-atoms, which can subsequently be deposited on a surface in a disordered fashion, or else on a pre-patterned template, to make ordered or periodic arrangements, for instance using capillary force assembly [53]. The possibility of dissociating these two scales (resonator/2D or 3D material) is quite unique to the bottom-up scheme.

Acknowledgments.

Final version published in EPL (Europhysics Letters) 119.1 (2017), p. 14004.

The authors acknowledge support from the LabEx AMADEus (ANR-10-LABX-42) in the framework of IdEx Bordeaux (ANR-10-IDEX-03-02), France.